\documentclass[aps,pre,twocolumn,floats,amsmath,amssymb,showpacs,floatfix]{revtex4-1}
\usepackage{dcolumn}
\usepackage{subfigure}
\usepackage{psfrag,graphicx, epsfig,color}

\begin{document}
\title{Prevalence of the sling effect for enhancing collision rates in turbulent suspensions}
\author{Michel Vo\ss kuhle, Alain Pumir, Emmanuel L\'ev\^eque}
\affiliation{Laboratoire de Physique,
 Ecole Normale Sup\'erieure de Lyon, CNRS, Universit\'e de Lyon,
 F-69007, Lyon, France}
\author{ Michael Wilkinson}
\affiliation{Department of Mathematics and Statistics,
The Open University, Walton Hall, Milton Keynes, MK7 6AA, England}

\begin{abstract}
Turbulence facilitates collisions between 
particles suspended in a turbulent flow. Two effects have
been proposed which can enhance the collision rate at
high turbulence intensities: \lq preferential concentration'
(a clustering phenomenon) and the \lq sling effect' (arising
from the formation of caustic folds in the phase-space of the 
suspended particles). 
We have determined numerically the collision 
rate of small heavy particles 
as a function of their size and densities.
The dependence on particle densities reveals that the enhancement by 
turbulence
of the collision rate of particles with significant inertia is due 
almost entirely to the sling effect. 
\end{abstract}

\maketitle

Understanding the rate of collisions between small particles, suspended in 
a turbulent fluid, is necessary for 
describing a variety of important physical processes.
In the case of clouds, collisions between droplets may determine the onset 
of rainfall~\cite{Shaw03}.
Models for planet formation involve aggregation through collisions of 
dust grains in the 
circumstellar disc~\cite{Safranov:69}. Last, collisions between suspended 
particles may 
be an important contribution
to dissipation of energy in some particle laden flows~\cite{Elghobashi:94}.
It is, therefore, 
of considerable
importance to quantify collisions between particles suspended in a turbulent
gas. 

The topic of collisions of particles suspended in a turbulent flow
has a long history, starting from the seminal work by Saffman and 
Turner, who were interested in understanding rain initiation in 
turbulent clouds~\cite{ST56}. Important theoretical insights have 
emerged in recent years, which indicate that the results of~\cite{ST56}
lead to an underprediction of the collision rate when the turbulence 
intensity is high, as a result of
two different mechanisms.
First, it has been shown that
particles can cluster due to an effect termed \lq preferential concentration',
which is ascribed to (heavy) particles being expelled from vortices by a centrifugal 
effect \cite{Max87} (other interpretations are considered in 
\cite{Wil+07}). This clustering effect can enhance the collision rate.
Second, it has been recognized that particles with inertia, which do 
{\it not} exactly follow the fluid motion, can both be arbitrarily close, and
yet have very different velocities. This effect induces collisions which 
may be thought as resulting from particles being \lq slung'  by 
vortices~\cite{FFS02}.
This phenomenon can also be understood in terms of caustics in the 
phase-space of the suspended particles~\cite{WM05,WMB06}. When the 
turbulence intensity is sufficiently high, a gas-kinetic model can be used
to describe the trajectories \cite{Abr75}, sometimes 
referred to as \lq random uncorrelated motion'~\cite{Meneguz:11}.
 
These mechanisms for enhancement of the collision rate have been illustrated by
simulations on model flows 
\cite{Meneguz:11,DP09}.
There have also been investigations of the collision
rates in simulations of fully-developed turbulence, 
which have  
provided quantitative information on preferential 
concentration and on the increase of the collision 
velocity~\cite{Sundaram:97,Wang:00,Rosa:13}.  
In this paper we report direct numerical simulation (DNS) studies of the collision 
rate of particles in fully-developed three-dimensional turbulence, as a function
of both their size and density. This extended parameter space
allows us to separate the clustering and the caustics/sling effect.  
We find that the caustics/sling effect is the dominant mechanism leading to the enhanced
collision rate in turbulent flows, 
even when the effect of particle inertia is expected to be relatively weak.

In the following paragraphs we discuss the clustering and 
sling/caustics 
models for the collision rate, before considering how these compare with our
numerical results. We consider a monodisperse suspension of spherical 
particles, of radius $a$, made of material
with density $\rho_{\rm p}$, suspended in an incompressible fluid
of density $\rho_{\rm f}$ and kinematic viscosity $\nu$. 
The fluid, with velocity field 
$\mbox{\boldmath$u$}(\mbox{\boldmath$r$},t)$, is in a statistically steady 
state of turbulent motion with rate of dissipation per unit mass equal to 
$\epsilon$.
We consider a sufficiently dilute suspension, so the flow is not significantly
perturbed by the presence 
of the particles. We assume that the particles obey the simple equation
of motion~\cite{MaxRil83,Gat83}:
\begin{equation}
\dot{\mbox{\boldmath$r$}}=\mbox{\boldmath$v$}
\ ,\ \ \ 
\dot{\mbox{\boldmath$v$}}=\frac{1}{\tau_{\rm p}}[\mbox{\boldmath$u$}(\mbox{\boldmath$r$},t)-\mbox{\boldmath$v$}]
\label{eq:MaxRil}
\end{equation}
where 
\begin{equation}
\tau_{\rm p}=\frac{2}{9}\frac{a^2}{\nu}\frac{\rho_{\rm p}}{\rho_{\rm f}}
\label{eq:tau_P}
\end{equation}
is the particle relaxation time, determined from Stokes formula for the 
drag on a moving sphere. This equation of motion is valid in the limit where 
the suspended particles are very small and very dense: 
$\rho_{\rm p}/\rho_{\rm f} \gg 1$.

In determining the motion of particles by using \eqref{eq:MaxRil}, only one
parameter is needed, namely the relaxation time $\tau_{\rm p}$. 
This time scale should be compared to a time scale of the flow.
To this end,
we introduce the Stokes number, as the ratio between $\tau_{\rm p}$ and  
the characteristic time of the flow
at the smallest scale, the  
Kolmogorov time scale $\tau_{\rm K} \equiv (\nu/\epsilon)^{1/2}$:
\begin{equation}
{\rm St}=\frac{\tau_{\rm p}}{\tau_{\rm K}}
\ .
\label{eq:def_St}
\end{equation}
The Stokes number parametrizes the effect of particle inertia.
For ${\rm St}\ll 1$, particles are advected by the fluid, and collisions are
the result of shear. When ${\rm St}\gg 1$, the inertia of the particles allows
them to move relative to the surrounding fluid. Note that ${\rm St}\propto \sqrt{\epsilon}$,
so that the inertial effects become more important when the 
turbulent intensity increases. In the range of droplet size most relevant
in the cloud microphysics context, $ 10 \mu {\rm m} \lesssim a \lesssim 20 \mu {\rm m}$, 
the Stokes number reaches at most values of the order of 
${\rm St} \sim 0.3$~\cite{Grab:99}. In other applications, such as planet formation  \cite{WMU08}, very large Stokes numbers are relevant.

We count a collision as occuring when the separation of the centers of 
independently moving particles come within $2a$.
The collision rate $R$, defined as the probability per unit time for a 
given particle to collide with any of the other particles, is
proportional to the volume density of the other particles, $n_0$, to 
the cross sectional 
area ($\propto a^2$) and to some appropriate average of the relative velocity for colliding 
particles, denoted by $\langle w\rangle$:
\begin{equation}
R = 4 \pi n_0 (2 a)^2 \langle w \rangle
\ .
\label{eq:def_R}
\end{equation}
The expected total number of collisions in a closed system of volume $V$ 
is simply obtained by multiplying $R$ by $n_0 V/2$. 
We neglected the role of gravity, and of hydrodynamic interactions
which may inhibit collisions by trapping a lubricating layer between the 
particles. We 
are concerned here with the collision rate for this slightly simplified model.
The objective is to describe
the collision rate determined from our DNS studies within 
the framework of a parametrisation based upon recent theoretical insights.

In the limit ${\rm St} \ll 1$ the collision rate is determined by shearing 
motion, so that $\langle w\rangle\sim (2 a)/\tau_{\rm K}$. Saffman and Turner 
argued that
\begin{equation}
R_{\rm ST}=\sqrt{\frac{8 \pi}{15} } \frac{n_0 (2 a)^3}{\tau_{\rm K}}
\ .
\label{eq:ST}
\end{equation}
Their calculation
includes all instances in which the separation radius decreases past 
$2a$. In the case of collisions where particles stick or coalesce on contact,
we should only count the first contact collisions. This effect should
be accounted for by introducing a factor $f<1$ in (\ref{eq:ST}). We ignore 
these corrections which will be discussed in a future publication, and 
simply set $f = 1$ here.

The enhancement of the collision rate, 
compared to the prediction of \eqref{eq:ST}, is expected to come 
from the particle trajectories breaking away from the fluid as the Stokes 
number increases. The effect 
termed preferential concentration causes clustering of particles with finite
values  of ${\rm St}$. The density of particles
at a distance $r$ from a given test particle is $n_0 g(r)$, where 
$g(r)$ is a radial correlation function. 
The relative 
velocity 
is not affected by the preferential concentration effect, so the collision rate
due to particles being advected into contact by shearing motion is
\begin{equation}
\label{eq:R_adv}
R_{\rm adv}= \sqrt{ \frac{8 \pi}{15} }\frac{n_0 (2 a)^3}{\tau_{\rm K}} g(2a)
\ .
\end{equation}
At a fixed Stokes number, the function $g (r)$ has a power-law dependence upon $r$:
$g(r) \propto r^\zeta$~\cite{RC:00}. 
This reflects the expectation that the suspended 
particles 
should sample a fractal measure \cite{Som+93,Bec03}. 
The exponent is $\zeta=d-D_2$,
where $D_2$ is the correlation dimension \cite{Gra+84}. DNS results 
indicate that for three dimensional turbulent flows, 
$2.3\le D_2 \le 3$ \cite{Bec+07}.  

In the limiting case where the turbulence intensity is very high, 
an alternative approach to understanding the effect of increasing the 
turbulence intensity was initiated by Abrahamson \cite{Abr75} , who pointed 
out that a gas-kinetic approach can be used 
to model the motion of the suspended particles. In this limit the relative 
velocity due to shearing motion induced by turbulence,
which is of order $a/\tau_{\rm K}$~\cite{ST56}, is replaced by a much 
larger relative 
velocity which characterises the relative motion of the fluid at different 
positions. This relative velocity may be parametrised by writing 
$\langle w\rangle\sim u_{\rm K}F({\rm St},{\rm Re})$, 
where $u_{\rm K} = (\epsilon \nu)^{1/4}  $ is the velocity at the 
Kolmogorov scale, $F$  depends on the Stokes number, and the Reynolds number,
${\rm Re}$. The collision rate 
is, therefore,
\begin{equation}
R_{\rm sling}=\frac{n_0a^2\eta}{\tau_{\rm K}}F({\rm St}, {\rm Re})
\ .
\label{eq:R_sling}
\end{equation}
The collision rate is the sum of contributions from collisions 
between particles which lie on the same branch of the phase-space 
manifold, giving rise to $R_{\rm adv}$, and collisions between
particles on different branches, giving rise to $R_{\rm sling}$:
\begin{equation}
\label{eq:R_sum}
R=R_{\rm adv}+ R_{\rm sling}
\end{equation}
This decomposition, proposed in earlier 
works~\cite{WMB06, DP09,Gustavsson:11}, rests on the assumption that the 
fraction of particles which give rise to preferential concentration, collide
with a small relative velocity with respect to the fluid, whereas another 
fraction, evenly distributed in the fluid, moves with large relative velocity.
The collisions due to these particles is described by the term $R_{\rm sling}$,
with the analytic form in \eqref{eq:R_sling}. When ${\rm St}\to 0$, the collision rate is well approximated 
by (\ref{eq:ST}), but
both terms can contribute to an enhanced collision rate as ${\rm St}$ increases.
The principal question addressed by this paper is to determine which 
contribution dominates as ${\rm St}$ increases.

It is possible to consider the asymptotic forms for the function 
$F({\rm St},{\rm Re})$
in equation (\ref{eq:R_sling}), in the limits of small and large Stokes numbers. In the limit as 
${\rm St}\to 0$, 
we must have $F({\rm St},{\rm Re})\to 0$, so that the limiting case (\ref{eq:ST}) 
is recovered from (\ref{eq:R_sum}). Considerations of model systems 
(described in \cite{WMB06}) suggest that $F$ has
 non-analytic behaviour in this limit, such 
as $F({\rm St},{\rm Re})\sim \exp(-C/{\rm St})$,
for some constant $C$: this is consistent with numerical results with the
Navier-Stokes equations~\cite{FP07}. 
The asymptotic form of the function $F({\rm St},{\rm Re})$ at large 
Stokes numbers has been considered by several authors. Abrahamson's theory is not valid for fully-developed turbulence, because it ignores the multiscale structure of the flow. 
A version which correctly accounts 
for the multiscale structure of turbulence was proposed by V\"olk {\sl et al.} \cite{Vol+80},
using the Kolmogorov model for the structure of the flow. This theory suggests
that $F({\rm St},\infty)\sim {\rm St}^{1/2}$. A simpler 
and more general argument was proposed in~\cite{Mehlig:07}: 
in the 
inertial range, the relative velocity can only depend upon $\epsilon $
and $\tau_{\rm p}$, so that dimensional analysis mandates that
$\langle w\rangle \sim \sqrt{\epsilon \tau_{\rm p}}$. Substituting for $\tau_{\rm p}$, we
have a rate of collision at high Stokes number which is of the form 
(\ref{eq:R_sling}) with 
$F({\rm St},\infty)\sim K\sqrt{{\rm St}}$, where $K$ is a universal dimensionless constant.
We emphasise that, because the preferential concentration effect
is a consequence of nearby particles experiencing a corrrelated strain-rate, 
this effect makes no contribution to $R_{\rm sling}$. 
Equation (\ref{eq:R_sling}) accounts for collisions 
between particles which have not experienced the same local environment, and 
the factor $g(2a)$ which occurs in (\ref{eq:R_adv}) is therefore absent from 
(\ref{eq:R_sling}).

\begin{figure}[t]
\begin{center}
\subfigure[]{
\includegraphics[width=0.40\textwidth]{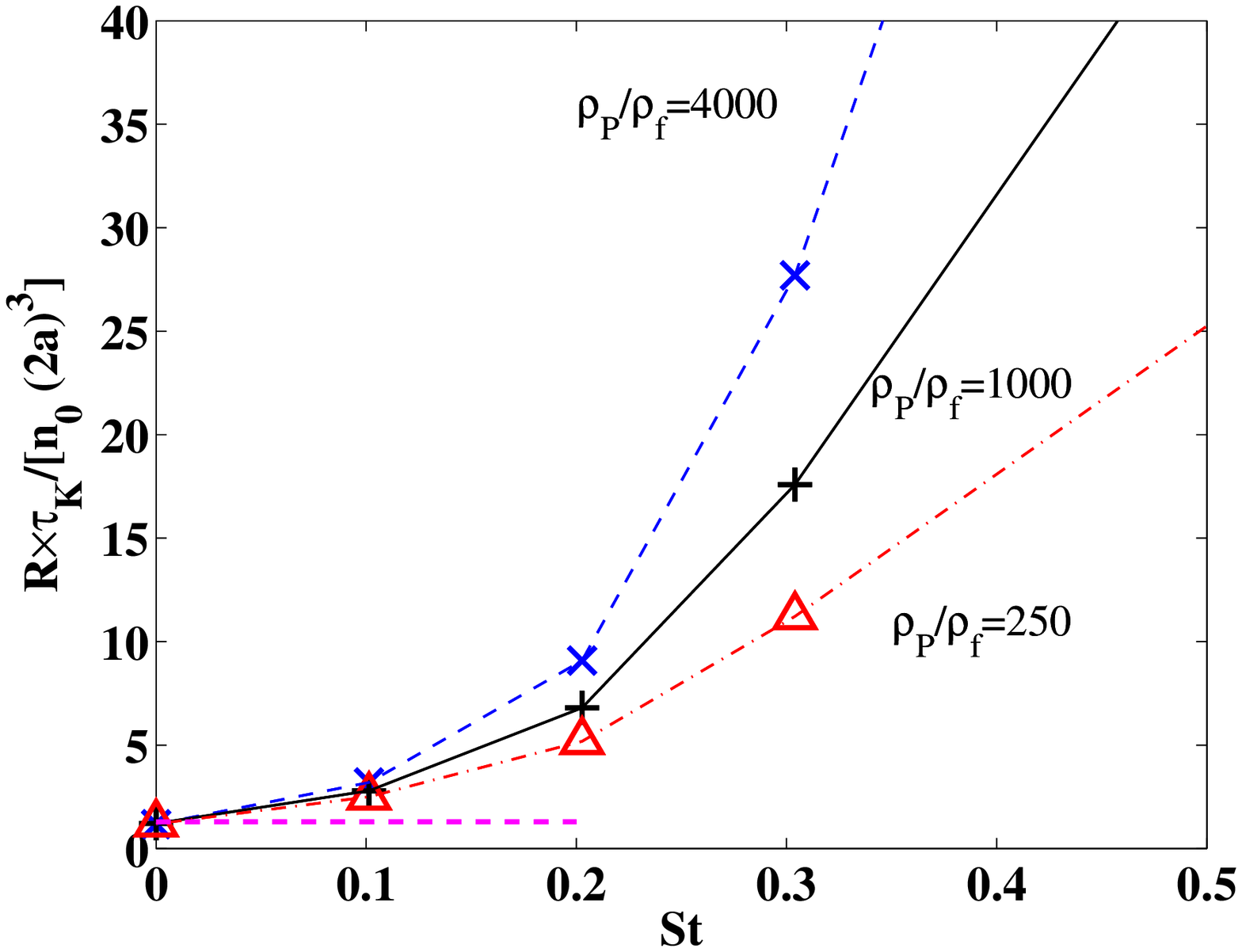}
}
\subfigure[]{
\includegraphics[width=0.40\textwidth]{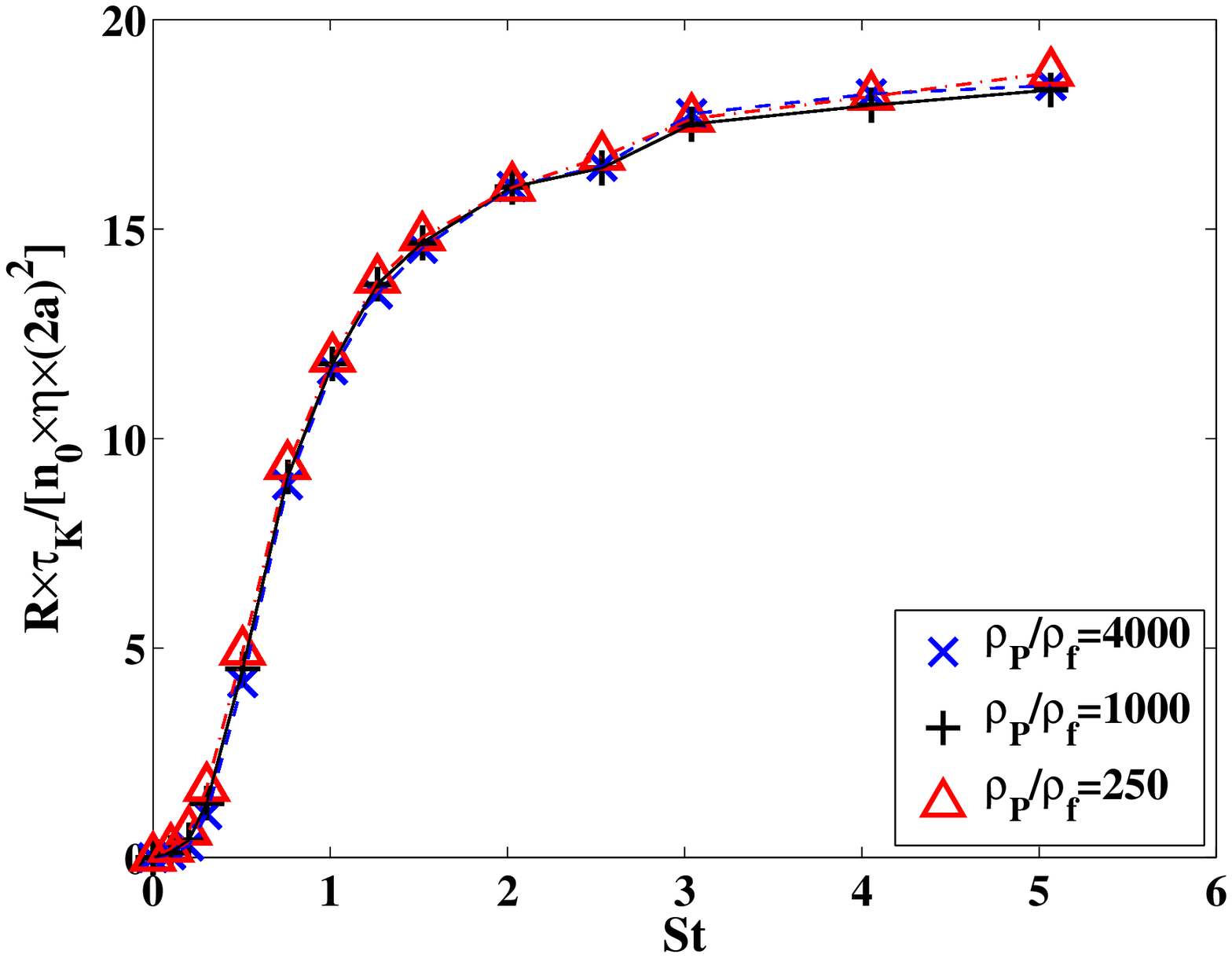}
}
\caption{\label{fig:coll_rate}The collision rate $R$ as a function of the Stokes number 
${\rm St}$ and for the ratios of density 
$\rho_{\rm p}/\rho_{\rm f} = 250$, $10^3$ and $4. 10^3$.
The collision rate $R$ is normalized by 
$n_0 (2a)^3/\tau_{\rm K}$ (a), and
$n_0  (2a)^2 \eta/\tau_{\rm K}$ (b).
The horizontal dashed line in (a) corresponds to the Saffman-Turner prediction.}
\end{center}
\end{figure}

We investigated the collision rate $R$ as a function of both $a$ and $\rho_{\rm p}/\rho_{\rm f}$.
Our simulations used a pseudo-spectral code, fully dealiased, with grid size $384^3$. The 
flow is forced  with a prescribed energy injection rate $\epsilon$~\cite{Lamorgese:05}.
The Taylor microscale
Reynolds number achieved in the steady state is ${\rm Re}_\lambda = 130$. 
Proper spatial resolution has been 
maintained, as can be judged from the product $k_{\rm max} \eta = 2$, where
$\eta = (\nu^3/\epsilon)^{1/4}$ is the  Kolmogorov scale, and $k_{\rm max}$
the largest wavenumber faithfully simulated. 
Particle trajectories were integrated by using the Velocity Verlet 
algorithm~\cite{NumRec} and resorting to tri-cubic interpolation to evaluate the fluid velocity at the position of the particle.
We detected collisions by using the algorithm described 
in ~\cite{Sundaram:96}.
Modifying the ratio $\rho_{\rm p}/\rho_{\rm f}$ at fixed value of
the Stokes number is achieved by varying 
in the collision detection algorithm the radius of the
particles, $a$, according
to \eqref{eq:tau_P},\eqref{eq:def_St} (so that
$a\propto (\rho_{\rm p}/\rho_{\rm f})^{-1/2}$). 
In the range of parameters considered, 
$\rho_{\rm p }/\rho_{\rm f} > 250$ and ${\rm St} \le 5$, the particle radii 
are at
most $\approx \eta /3$, which ensures that \eqref{eq:MaxRil} provides a very
good description of the motion. 
We find that after a transient state of $\approx 5$
eddy turnover times,
the collision rate becomes independent of time.
The collision rates were determined by recording
at the minimum $1.3 \times 10^4$ collisions 
when $\rho_{\rm p}/\rho_{\rm f} = 10^3$,
accumulated (except in one case) 
over $\gtrsim 10$ eddy-turnover times.

The collision rate, $R$, determined numerically is plotted in 
Fig.~\ref{fig:coll_rate}. As explained earlier, we do not distinguish here
between single and multiple collisions. 
In Fig.~\ref{fig:coll_rate}(a), $R$ is normalized by
$n_0 (2 a)^2/\tau_{\rm K}$ and plotted as a function of ${\rm St}$.
The Saffman-Turner prediction, \eqref{eq:ST}, implies that in the limit 
$St\rightarrow 0$, the quantity $R \tau_{\rm K}/(n_0 (2a)^3)$ should become 
independent
of the ratio $\rho_{\rm p}/\rho_{\rm f}$. Our own numerical results are
only consistent with this prediction for small values of ${\rm St}$. 
Fig.~\ref{fig:coll_rate}(b) shows
that $R \tau_{\rm K}/(n_0a^2\eta)$
as a function of the Stokes number, 
does not depend much on $\rho_{\rm p}/\rho_{\rm f}$ for values of ${\rm St}$
larger than $\gtrsim 0.3$.
This scaling is consistent with the sling/caustics 
collision mechanism, described by equation \eqref{eq:R_sling}. 
We note that $F(St, Re)$ deduced from Fig.~\ref{fig:coll_rate}(b) does not 
fit the asymptotic 
form $F({\rm St},\infty)=K\sqrt{{\rm St}}$ for large values of ${\rm St}$. We ascribe
this to the limited Reynolds number of our numerical simulations.

\begin{figure}[t]
\includegraphics[width=0.40\textwidth]{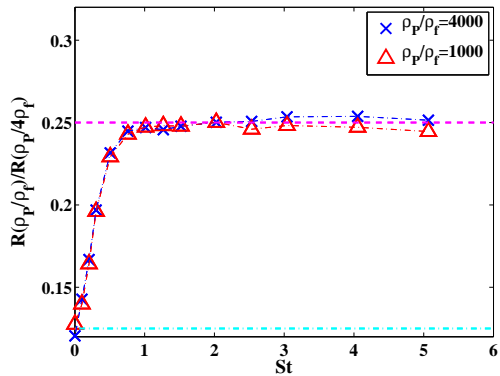}
\caption{
The ratio between the collision rates $R$ corresponding to 
$\rho_{\rm p}/\rho_{\rm f} = 4000$ and $1000$ (crosses) and 
$\rho_{\rm p}/\rho_{\rm f} = 1000$ and $250$ (triangles), 
illustrating the crossover between the sling dominated
regime for ${\rm St} \gtrsim 0.5$, and the regime described by the 
Saffman-Turner theory for ${\rm St} \lesssim 0.2$. 
\label{fig:coll_rate_ratio}}
\end{figure}

A clear illustration of the transition from the regime described by the 
Saffman-Turner prediction, \eqref{eq:ST}, and the sling dominated regime, 
\eqref{eq:R_sling}, is provided by  
Fig.~\ref{fig:coll_rate_ratio} which shows the ratio between the values
of $R$ computed at $\rho_{\rm p}/\rho_{\rm f} = 4000$ and $1000$ 
(crosses)
and $\rho_{\rm p}/\rho_{\rm f} = 1000$ and $250$ (triangles). Whereas 
\eqref{eq:ST}
predicts that these ratios should be $1/8$, \eqref{eq:R_sling} predicts
rather a ratio of $1/4$. Fig.~\ref{fig:coll_rate_ratio} shows that the
ratios are extremely close to $1/4$ for $St \gtrsim 0.5$, but 
approaches $1/8$ for ${\rm St} \lesssim 0.3$. 

\begin{figure}[t]
\includegraphics[width=0.40\textwidth]{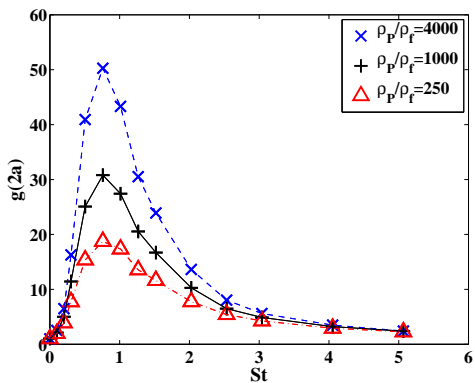}
\caption{The function $g$ that measures preferential concentration, 
computed for three values of $a$, corresponding to particles with a density
$\rho_{\rm p}$ equal to $250 \rho_{\rm f}$, $1000 \rho_{\rm f}$ and 
$4000 \rho_{\rm f}$, as indicated in the figures.
The preferential concentration does not play a significant role for 
${\rm St} \gtrsim 5$ 
\label{fig:pref_conc}
}
\end{figure}

\begin{figure}[t]
\includegraphics[width=0.40\textwidth]{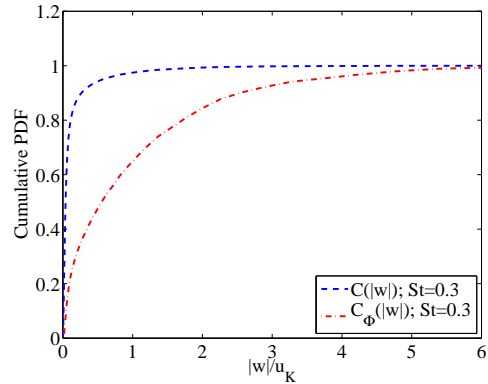}
\caption{The cumulative distribution of radial velocities of colliding 
particles, $C(|w|)$ (dashed line), and the cumulative distribution weighted by 
$|w|$, $C_{\Phi}( |w| )$ (dashed-dotted line), defined by Eq.~(\ref{eq: 9}).
These describe the 
contribution to the collision rate due to 
particle pairs colliding with relative velocity less than $|w|$. While particles
with a velocity larger than $u_{\rm K}$ are very few, they are responsible for a
sizable fraction of the collision rate. The data shown corresponds to 
${\rm St} = 0.3$; $u_{\rm K} \times \tau_{\rm K}/(2a) \approx 27$, and $\rho_{\rm p}/\rho_{\rm f} = 1000$. 
\label{fig:cum_PDF}
}
\end{figure}

Fig.~\ref{fig:pref_conc} shows the function $g(2a)$, (which
quantifies the importance of preferential concentration) 
in our simulations for the three different values of 
$\rho_{\rm p}/\rho_{\rm f}$. 
The strong enhancement of the concentration at the surface of 
a particle is not  sufficient to make the advective collision rate 
(\ref{eq:R_adv})
comparable to the sling collision rate, (\ref{eq:R_sling}).

Further evidence for the importance of caustics comes from
considering the probability density, $P(w|2a)$, of the radial relative velocity between two 
particles,
$w \equiv \delta {\mathbf{v}} \cdot \delta {\mathbf{r}}/|\delta {\mathbf{r}}|$, 
conditioned on the fact that the two particles collide ( 
$ | \delta \mathbf{r} | = 2a $ and $w \le 0$).
Fig.~\ref{fig:cum_PDF} shows the cumulative PDF, $C(|w|)$, and 
the contribution of particles of velocity $w' < |w|$ to the flux~\cite{Sundaram:97}, 
$C_\Phi(|w|)$:
\begin{eqnarray}
\label{eq: 9}
C(|w|)& =& \int_0^{|w|} P(w'|2a)\ {\rm d}w'\nonumber \\
C_\Phi(|w|) &=&\frac{ \int_0^{|w|} P(w'|2a)\ w'\ {\rm d}w'}{\int_0^{\infty} P(w' | 2a)\ w'\ {\rm d}w'}
\end{eqnarray}
Even for values of the Stokes number as low as ${\rm St} = 0.3$, $\sim 90\%$ of all
particle pairs have a relative velocity difference less than 
$ |w| \lesssim 8 (2a)/\tau_{\rm K}$, 
but contribute only to $\sim 37\%$ of 
the collision rate. Fig.~\ref{fig:cum_PDF} therefore demonstrates that
the contribution of the sling term is the prevalent effect responsible for
the large increase of the collision rate, even at moderate Stokes numbers.

An alternative decomposition, originally proposed in~\cite{Sundaram:97}, 
expresses
the collision rate $R$ as a product in which the term $g(2a)$, which 
describes the
local concentration enhancement around a particle, appears as 
an overall factor:
\begin{equation}
\label{eq: 10}
 R = 4 \pi (2a)^2 g(2a) \langle w \rangle_{\rm eff}
\end{equation}
This representation, which is exact for a suitable definition 
of $\langle w \rangle_{\rm eff}$, suggests that 
the preferential concentration and sling effects act together to 
enhance the collision rate. Figure \ref{fig:coll_rate}(b)
demonstrates that if this parametrisation of the collision
rate is used, then the dependence of $g(2a)$ upon $\rho_{\rm p}/\rho_{\rm f}$
shown in figure \ref{fig:pref_conc} must be cancelled 
(for ${\rm St}\ge 0.5$) by a reciprocal dependence of the collision velocity,
$\langle w \rangle_{\rm eff}$.  
In fact,  
previous measurements~\cite{Bec:10,Rosa:13} of the dependence of $g(r)$ 
and of the average velocity difference as a function of $r$ suggest power law
dependences, the exponents being such that the product 
$g(2a) \langle w \rangle_{\rm eff}$
is essentially constant for ${\rm St } \gtrsim 0.5$.
Our equations 
(\ref{eq:R_adv}), (\ref{eq:R_sling}) and (\ref{eq:R_sum}) give 
a physically well-motivated theory which explains the data, and provide
an explanation for this cancellation.

We conclude that in turbulent flows and at 
large values of $\rho_{\rm p}/\rho_{\rm f}$
(the case relevant to typical aerosols), 
the sling effect provides the dominant mechanism for the dramatically
enhanced collision rate, for particles whose Stokes number exceeds $\sim 0.3$. 

The authors acknowledge informative discussions with B. Mehlig, 
K. Gustavsson and L. Collins. AP and EL have been supported by the grant from 
A.N.R. \lq TEC 2'. Computations were performed at the PSMN computing center
at the Ecole Normale Sup\'erieure de Lyon.
MW and AP were supported by the EU COST action MP0806 \lq Particles in Turbulence'.


\begin{thebibliography}{}

\bibitem{Shaw03}
R. A. Shaw, 
{\it Annu. Rev. Fluid Mech.} {\bf 35}, 183-227 (2003).

\bibitem{Safranov:69}
V. S. Safranov, 
{\sl Evolution of the protoplanetary cloud and formation of earth and planets},
{\it NASA Tech. Transl. F-677; Moscow, Nauka} (1969).


\bibitem{Elghobashi:94} 
S. Elghobashi,
{\it Applied Scientific Research} {\bf 52}, 309 (1994).


\bibitem{ST56}
P. G. Saffman and J. S. Turner,
{ \it J. Fluid. Mech. } {\bf 1}, 16-30 (1956).

\bibitem{Max87}
M. R. Maxey, 
{\it J. Fluid Mech.}, {\bf 174}, 441, (1987).

\bibitem{Wil+07}
M. Wilkinson, B. Mehlig, S. \"Ostlund and K. P. Duncan,
{\it Phys. Fluids}, {\bf 19}, 113303, (2007).

\bibitem{FFS02}
G. Falkovich, A. Fouxon and M. G. Stepanov,
{ \it Nature } {\bf 419}, 151-154 (2002).

\bibitem{WM05}
M. Wilkinson and B. Mehlig,
{ \it Europhys. Lett. } {\bf 71}, 186 (2005).

\bibitem{WMB06}
M. Wilkinson, B. Mehlig and V. Bezuglyy,
{ \it Phys. Rev. Lett. } {\bf 97}, 048501 (2006).

\bibitem{Abr75}
J. Abrahamson,
{\it Chem. Eng. Sci.} {\bf 30}, 1371-1379 (1975).

\bibitem{Meneguz:11}
E. Meneguz and M. W. Reeks,
{\it J. Fluid Mech.} {\bf 686}, 338-351 (2011).

\bibitem{DP09}
L. Ducasse and A. Pumir,
{\it Phys. Rev.} {\bf E 80}, 066312 (2009).

\bibitem{Sundaram:97}
S. Sundaram and L. R. Collins,
{\it J. Fluid Mech. }{\bf 335}, 75 (1997).

\bibitem{Wang:00}
L.P. Wang, A. S. Wexler and Y. Zhou,
{\it J. Fluid Mech. }{\bf 415}, 117 (2000).


\bibitem{Rosa:13}
B. Rosa, H. Parishani, O. Ayala, W. W. Grabowski and L. P. Wang,
{\it New J. Phys.} {\bf 15}, 045032 (2013).


\bibitem{MaxRil83}
M. R. Maxey and J. J. Riley,
{\it Phys. Fluids} {\bf 26}, 883 (1983)

\bibitem{Gat83}
R. Gatignol,
{\it J. Mec. Theor. Appl.}, {\bf 1}, 143 (1983).

\bibitem{Grab:99}
W. Grabowski and P. Vaillancourt, 
{\it J. Atmos. Sci.}, {\bf 56}, 1433-1436 (1999).

\bibitem{WMU08}
M. Wilkinson, B. Mehlig and V. Uski,
{\it Astrophys. J., Suppl.},  {\bf 176}, 484, (2008).

\bibitem{RC:00} W. C. Reade and L. R. Collins,
{\it Phys. Fluids }, {\bf 12}, 2530, (2000). 

\bibitem{Som+93}
J. Sommerer and E. Ott,
{\it Science}, {\bf 259}, 335, (1993).

\bibitem{Bec03}
J. Bec,
{\it Phys. Fluids}, {\bf 15}, L81-4, (2003).

\bibitem{Gra+84}
P. Grassberger and I. Procaccia, 
{\it Physica D}, {\bf 9}, 189-208,  (1983).

\bibitem{Bec+07}
J. Bec, L. Biferale, M. Cencini, A. Lanotte, S. Musacchio, and F. Toschi, 
{\it Phys. Rev. Lett.}, {\bf 98}, 084502, (2007).


\bibitem{Gustavsson:11}
K. Gustavsson and B. Mehlig,
{\it Phys. Rev. } {\bf E 84}, 045304 (2011).

\bibitem{FP07}
G. Falkovich and A. Pumir,
{ \it J. Atmos. Sci. } {\bf 64}, 4497-4505 (2007).

\bibitem{Vol+80}
H. J. V\" olk, F. C. Jones, G. E. Morfill, and S. R\"oser, 
{\it Astron. Atrophys. }, {\bf 85}, 316, (1980).

\bibitem{Mehlig:07}
B. Mehlig, V. Uski and M. Wilkinson,
{\it Phys. Fluids } {\bf 19}, 098107 (2007).


%
%

\bibitem{Lamorgese:05}
A. G. Lamorgese, D. A. Caughey and S. B. Pope,
{\it Phys. Fluids} {\bf 17}, 015106 (2005).

\bibitem{NumRec}
W. H. Press, S. A. Teukolsky, W. T. Vetterling and B. P. Flannery,
{\it Numerical Recipes: the art of scientific computing}, New York, Cambridge
University Press (2007).

\bibitem{Sundaram:96}
S. Sundaram and L. R. Collins,
{\it J. Comput. Phys. }{\bf 124}, 337 (1996).


\bibitem{Bec:10}
J. Bec, L. Biferale, M. Cencini, A. S. Lanotte and F. Toschi,
{\it J. Fluid Mech.} {\bf 646}, 527-536 (2010).


\end{thebibliography}
\end{document}